\documentclass{article}

\usepackage{cite}
\usepackage{natbib}
\usepackage[colorlinks=true, allcolors=blue]{hyperref}
\usepackage{amsthm}
\usepackage{amsmath, bm, amssymb}
\usepackage{graphicx}
\usepackage{subcaption}
\usepackage{adjustbox}
\usepackage{float}

\newcommand{\pkg}[1]{\textbf{#1}}           
\newcommand{\proglang}[1]{\textsf{#1}}      
\usepackage{booktabs}
\usepackage{longtable}
\usepackage{array}
\usepackage{multirow}
\usepackage{wrapfig}
\usepackage{colortbl}
\usepackage{pdflscape}
\usepackage{tabu}
\usepackage{threeparttablex}
\usepackage[normalem]{ulem}
\usepackage{makecell}
\usepackage{xcolor}
\usepackage{orcidlink,thumbpdf,lmodern}
\usepackage{algorithm, algcompatible, setspace}
\usepackage{booktabs}

\usepackage{authblk}

\usepackage[skip=10pt plus1pt]{parskip}
\usepackage{setspace}
\usepackage[letterpaper,top=2cm,bottom=2cm,left=3cm,right=3cm,marginparwidth=1.75cm]{geometry}
\title{\pkg{modelimportance}: An \proglang{R} package for evaluating model importance within a multi-model ensemble}
\author[1]{Minsu Kim\thanks{Corresponding author}}
\author[1]{Li Shandross}
\author[1]{Evan L. Ray}
\author[1]{Nicholas G. Reich}

\affil[1]{\small{\textit{School of Public Health and Health Sciences, University of Massachusetts Amherst, Amherst, MA 01003, USA}}}
\date{}

\begin{document}
\maketitle
\begin{abstract}
Ensemble forecasts are commonly used to support decision-making and policy planning across various fields because they often offer improved accuracy and stability compared to individual models. As each model has its own unique characteristics, understanding and measuring the value of each constituent model can support the construction of effective ensembles. 
The \proglang{R} package \pkg{modelimportance} provides tools to quantify how each component model contributes to the accuracy of ensemble performance for both point and probabilistic forecasts. The package supports multiple ensemble methods and multiple model importance metrics. 
Additionally, the software offers customizable options for handling missing values. These features enable the package to serve as a versatile tool for researchers and practitioners. It helps not only in constructing an effective ensemble model across a wide range of forecasting tasks, but also in understanding the role of each model within the ensemble and gaining insights into individual models themselves. This package follows the `hubverse' framework, which is a collection of open-source software, tools and data standards developed to promote collaborative modeling hub efforts and simplify their setup and operation. 
Doing so enables seamless integration and flexibility with other forecasting tools and systems, allowing many analyses to be performed on existing hubs.
\end{abstract}

\section{Introduction}\label{ch2sec:intro}
Ensemble forecasting is a method to produce a single, consolidated prediction by combining forecasts generated from different models. 
While individual models' strengths and weaknesses are pronounced within ensembles, the combination of many models tend to offset each other and create an ensemble forecast that is more robust and accurate than any single component model \cite{gneiting2005weather, hastie01statisticallearning}. 
Specifically, ensembles effectively mitigate the bias and variance arising from the predictions of individual models by averaging them out, and aggregating in this way can reduce prediction errors and improve overall performance. 
Enhanced prediction accuracy and robustness enable the achievement of more reliable predictions, thereby improving decision-making. For this reason, ensemble forecasting is widely used across various domains such as weather forecasting \cite{Guerra_2020, gneiting2005weather}, financial modeling \cite{SUN2020101160, math11041054}, and infectious disease outbreak forecasting \cite{ray_prediction_2018, reich_accuracy_2019, lutz_applying_2019, viboud_rapidd_2018} to name a few. 
For example, throughout the COVID-19 pandemic, the US COVID-19 Forecast Hub collected individual models developed by over 90 different research groups and built a probabilistic ensemble forecasting model for COVID-19 cases, hospitalizations, and deaths in the US, based on those models' predictions. The ensemble model served as the official short-term forecasts for the US Centers for Disease Control and Prevention (CDC) \cite{cramer2022united}.

The quality of forecasts is assessed by evaluating their errors, biases, sharpness, and/or calibration using different scoring metrics. 
The selection of scoring metrics depends on the type of forecast: point forecasts (e.g., mean, median) and probabilistic forecasts (e.g., quantiles, probability mass function). 
Commonly used assessment tools for point forecasts are the mean absolute error (MAE) and the mean squared error (MSE), which calculate the average magnitude of forecast errors. 
Scoring metrics for probabilistic forecasts consider the uncertainty and variability in predictions and provide concise evaluations through numerical scores \cite{gneiting_strictly_2007}.
Some examples include the weighted interval score (WIS), the continuous ranked probability score (CRPS), and the log score \cite{bracher_evaluating_2021}. (Note that CRPS is a general scoring rule that can be computed either analytically in closed form or numerically from samples, and WIS is a quantile-based approximation of CRPS.)

Several \proglang{R} packages have been developed to evaluate forecast quality across both point and probabilistic settings. 
To name a few, the \pkg{fable} package \cite{Rpackage-fable} is widely used for univariate time series forecasting and includes functions for accuracy measurement.
The \pkg{Metrics} \cite{Rpackage-Metrics} and \pkg{MLmetrics} \cite{Rpackage-MLmetrics} provide a wide range of performance metrics specifically designed for evaluating machine learning models. 
The \pkg{scoringRules} \cite{Rpackage-scoringRules} package offers a comprehensive set of proper scoring rules for evaluating probabilistic forecasts and supports both univariate and multivariate settings. 
The \pkg{scoringutils} \cite{bosse2022evaluating} package offers additional features to the functionality provided by \pkg{scoringRules}, which makes it more useful for certain tasks, such as summarizing, comparing, and visualizing forecast performance. These packages have been valuable to evaluate individual models as independent entities, using performance metrics selected for each specific situation or problem type. 
However, they do not measure the individual models' contributions to the enhanced predictive accuracy when used as part of an ensemble. 
Building on our prior methodological study \cite{kim2026beyond}, we emphasize that strong standalone model performance does not automatically imply a large positive contribution once the model is used within an ensemble. 
The \pkg{modelimportance} package operationalizes this idea in software, providing tools to evaluate the role of each model as an ensemble member within an ensemble model, rather than focusing on the individual predictive performance per se.

In ensemble forecasting, certain models contribute more significantly to the overall predictions than others. Assessing the impact of each component model on ensemble predictions is methodologically similar to determining variable importance in traditional regression and machine learning models, where variable importance measures evaluate how much individual variables improve the accuracy of the model's predictive performance or reduce the average loss. \proglang{R} packages such as \pkg{randomForest} \cite{Rpackage-randomForest}, \pkg{caret}\cite{Rpackage-caret}, \pkg{xgboost} \cite{Rpackage-xgboost}, and \pkg{gbm} \cite{Rpackage-gbm} implement these functions for different types of models: random forest models, general machine learning models, extreme gradient boosting models, and generalized boosted regression models, respectively. These packages focus on feature-level importance within a single model and do not measure the contribution of individual models within an ensemble. In contrast, the tools in \pkg{modelimportance} quantify how each component model helps enhance the ensemble model's predictive performance. 
They assign numerical scores to each model using a forecast accuracy metric selected based on the forecast type.

Our methods are based on the concept of Shapley values in cooperative game theory, which measure a player's average contribution to the game's overall outcome (\cite{Shapley1953}). 
There are several approaches that utilize Shapley values in black-box machine learning models to understand how each feature affects the model's predictive power (\cite{lundberg2017unified}; \cite{lundberg2020local}; \cite{covert2020understanding}). 
Related work by \cite{lipiecki2024postprocessing} used Shapley-based attribution to study contributions of component models in a combined ensemble after postprocessing point forecasts into probabilistic forecasts. 
That methodology was later implemented in \pkg{PostForecasts.jl} \cite{Jlpackage-lipiecki2025postforecast}. \pkg{PostForecasts.jl} is conceptually aligned with our work in that both approaches leverage Shapley values to evaluate component forecasters. However, while \pkg{PostForecasts.jl} was developed for \proglang{Julia} users originally focusing on energy economics, our \proglang{R} package focuses on epidemiological forecasts and offers a broader range of capabilities, such as handling missing forecasts, compatibility with the hubverse forecasting ecosystem, and rich tools for summarizing and visualizing importance scores across tasks.

These capabilities are particularly useful for hub organizers, who oversee collaborative forecasting systems that combine submissions from many teams into a single ensemble forecast \cite{shandross2026multi}.
Examples include efforts coordinated by the US CDC and the European Centre for Disease Prevention and Control. By quantifying each component model's contribution, \pkg{modelimportance} can support evidence-based decisions about ensemble design and maintenance. 
The package follows conventions defined by the `hubverse', a community-maintained ecosystem of open software and data standards for collaborative forecasting hubs \cite{hubverse_docs, kerr2025coordinating}. Using these shared data structures allows our package to be seamlessly integrated with existing hub pipelines and deployed across multiple active hubs. 
We note that there are 31 hubs as of March 2026, including model development and training hubs.

We highlight some development practices we employed, such as unit testing of individual functions, object-oriented programming using S3 classes, continuous integration testing on different operating systems, and independent code review by peer developers. This emphasis on quality control is a key strength of this work.

The paper proceeds as follows.
\hyperref[ch2sec:data]{Section~\ref{ch2sec:data}} describes (a) how the \pkg{modelimportance} package relates to the hubverse framework, including its dependencies, (b) the model output formats defined within hubverse, and (c) the structure of data presentation for both forecasts and actual observations.
\hyperref[ch2sec:algorithms]{Section~\ref{ch2sec:algorithms}} presents two algorithms implemented in \pkg{modelimportance} for calculating the model importance metric: leave-one-model-out and leave-all-subsets-of-models-out.
\hyperref[ch2sec:main-function]{Section~\ref{ch2sec:main-function}} demonstrates the various functionalities \pkg{modelimportance} supports.
\hyperref[ch2sec:s3-infrastructure]{Section~\ref{ch2sec:s3-infrastructure}} describes the S3 class structure and related methods implemented in the package, followed by examples of their usage in \hyperref[ch2sec:examples]{Section~\ref{ch2sec:examples}}.
\hyperref[ch2sec:computational-complexity]{Section~\ref{ch2sec:computational-complexity}} discusses computational complexity and strategies for efficient computation.
\hyperref[ch2sec:implementation-and-availability]{Section~\ref{ch2sec:implementation-and-availability}} highlights our quality assurance measures and the code's availability, and then we close this paper with a summary and a discussion of possible extensions.

\section{Data}\label{ch2sec:data}
\subsection{Dependencies and related software}\label{ch2subsec:dependences}

The \pkg{modelimportance} package is designed to work with the hubverse framework and, accordingly, depends on several packages in the hubverse ecosystem, such as \pkg{hubUtils} \cite{Rpackage-hubUtils}, \pkg{hubEnsembles} \cite{Rpackage-hubEnsembles}, and \pkg{hubEvals} \cite{Rpackage-hubEvals}.  
\pkg{modelimportance} uses a \texttt{model\_out\_tbl} S3 class as the model output format defined in \pkg{hubUtils}, which consists of utility functions to standardize prediction files and data formats (details in \hyperref[ch2subsec:model_output_format]{Section~\ref{ch2subsec:model_output_format}}).
Ensembling predictions from multiple models relies on \pkg{hubEnsembles}, which offers a broadly applicable framework to construct multi-model ensembles using various ensemble methods.
Calculation of forecast accuracy using various metrics is based on \pkg{hubEvals}, which internally leverages \pkg{scoringutils}. 

We derived example datasets used for testing and demonstration purposes (see \hyperref[ch2sec:examples]{Section~\ref{ch2sec:examples}}) from \pkg{hubExamples} package \cite{Rpackage-hubExamples}, which provides example datasets in the hubverse format. The dataset were locally stored 
for reproducibility within the package distribution, instead of requiring the \pkg{hubExamples} package as a dependency.

For visualization of model importance scores, standard \pkg{ggplot2} \cite{Rpackage-ggplot2} functions can be applied. Parallel computing is supported by the \pkg{furrr} \cite{Rpackage-furrr} and \pkg{future} \cite{Rpackage-future} packages, which allow users to specify the number of cores to use for parallel processing when calculating model importance scores across many tasks.

\subsection{Model output format}\label{ch2subsec:model_output_format}

Model outputs are structured in a tabular format designed specifically for predictions, which is a formal S3 object called \texttt{model\_out\_tbl}. In the hubverse standard, each row represents an individual prediction or a component of a prediction for a single task. More details about that prediction or prediction component are described in multiple columns through which one can identify the unique label assigned to each forecasting model, task characteristics, prediction representation type, and predicted values \cite{shandross2026multi}. 
To elaborate on the task characteristics, each prediction task corresponds to a specific forecasting problem and it can be described by a set of task ID variables. Examples of such variables include a date on which forecasts are generated, the target to predict (e.g., flu-related incident deaths, cases, or hospitalizations), and the prediction horizon, which is the length of time into the future from the point when a model generate its forecast, for a specific location on a certain target date. Table~\ref{tbl-example-model_output} illustrates short-term forecasts of weekly incident influenza hospitalizations in the US for Massachusetts, generated by the model `Flusight-baseline' on December 17, 2022, in the \texttt{model\_out\_tbl} format. 
The \texttt{model\_id} column lists a uniquely identified model name. The \texttt{reference\_date}, \texttt{target}, \texttt{horizon}, \texttt{location}, and \texttt{target\_end\_date} columns are collectively referred to as the task ID variables, which together defines the task characteristics. 
Note that the forecast generation date and the target date for which the prediction is made are mapped to the \texttt{reference\_date} and \texttt{target\_end\_date} columns, respectively, and the location is represented based on the FIPS code (e.g., `25' for Massachusetts). 
The time length to the \texttt{target\_end\_date}, which is the number of weeks ahead from the \texttt{reference\_date}, is indicated in the \texttt{horizon} column. The prediction representation is specified as `quantile' in the \texttt{output\_type} column, and details are represented in the \texttt{output\_type\_id} column with seven quantiles of 0.05, 0.1, 0.25, 0.5, 0.75, 0.9, and 0.95 for each target end date.
The predicted value corresponding to each quantile is recorded in the \texttt{value} column.

\begin{table}
\centering{
\centering
\resizebox{\ifdim\width>\linewidth\linewidth\else\width\fi}{!}{
\begin{tabular}{lllrllllr}
\toprule
model\_id & reference\_date & target & horizon & location & target\_end\_date & output\_type & output\_type\_id & value\\
\midrule
Flusight-baseline & 2022-12-17 & wk inc flu hosp & 1 & 25 & 2022-12-24 & quantile & 0.05 & 496\\
Flusight-baseline & 2022-12-17 & wk inc flu hosp & 1 & 25 & 2022-12-24 & quantile & 0.1 & 536\\
Flusight-baseline & 2022-12-17 & wk inc flu hosp & 1 & 25 & 2022-12-24 & quantile & 0.25 & 566\\
Flusight-baseline & 2022-12-17 & wk inc flu hosp & 1 & 25 & 2022-12-24 & quantile & 0.5 & 582\\
Flusight-baseline & 2022-12-17 & wk inc flu hosp & 1 & 25 & 2022-12-24 & quantile & 0.75 & 598\\
\addlinespace
Flusight-baseline & 2022-12-17 & wk inc flu hosp & 1 & 25 & 2022-12-24 & quantile & 0.9 & 629\\
Flusight-baseline & 2022-12-17 & wk inc flu hosp & 1 & 25 & 2022-12-24 & quantile & 0.95 & 668\\
Flusight-baseline & 2022-12-17 & wk inc flu hosp & 2 & 25 & 2022-12-31 & quantile & 0.05 & 454\\
Flusight-baseline & 2022-12-17 & wk inc flu hosp & 2 & 25 & 2022-12-31 & quantile & 0.1 & 518\\
Flusight-baseline & 2022-12-17 & wk inc flu hosp & 2 & 25 & 2022-12-31 & quantile & 0.25 & 558\\
\bottomrule
\end{tabular}}
}
\caption{\label{tbl-example-model_output}Example of the model output for incident influenza hospitalizations (top 10 rows) extracted from \texttt{forecast\_data\_example} data bundled in the \pkg{modelimportance} package, which is originally sourced from the \pkg{hubExamples} package.}
\end{table}%

\begin{figure}[t]
\centering{
    \includegraphics{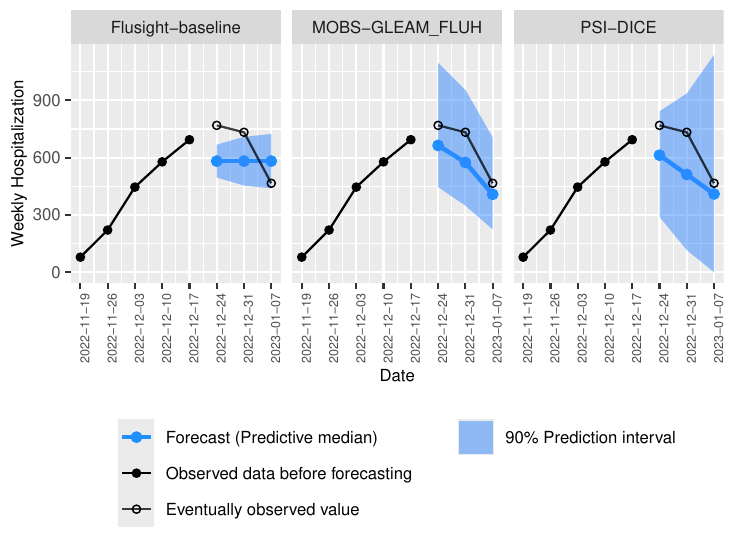}
}
\caption{\label{fig-example-model_output}Example plot of three distributional forecasts corresponding to the model output for incident influenza hospitalizations shown in Table~\ref{tbl-example-model_output}. Solid black dots indicate historically available data as of the forecast generation date, and open black circles indicate the eventually observed values. The blue dots represent predictive medians and the blue shaded area represents the corresponding 90\% prediction interval defined by the 0.05 and 0.95 quantiles.} 
\end{figure}

Figure~\ref{fig-example-model_output} visualizes the information on the prediction task provided by Table~\ref{tbl-example-model_output} for three models. For each model, the quantile-based forecasts are shown for the target end dates of December 24, 2022 (horizon 1), December 31, 2022 (horizon 2), and January 07, 2023 (horizon 3), which were made on December 17, 2022 based on the historical data available as of that date. 
The prediction intervals defined by the lowest and highest quantiles (0.05 and 0.95) represent the uncertainty of the predictions.
To give a brief interpretation, the Flusight-baseline model under-predicted the outcomes for the first two target dates (horizon 1 and 2), but it over-predicted the outcome for the last target date (horizon 3). 
Its prediction intervals are narrow compared to the other two models, which indicates that it is more confident about its predictions. However, two of three prediction intervals (horizons 1 and 2) failed to cover the eventually observed values, implying that the model was overconfident and/or biased.

\subsection{Forecast data representation}\label{ch2subsec:model_output}

Generally, quantitative forecasts can be categorized as being either point forecasts or probabilistic forecasts. 
For a specific prediction task, point forecasts, represented by a single predicted value, provide a clear and concise prediction, making them easy to interpret and communicate. 
Probabilistic forecasts, on the other hand, provide a probability distribution over possible future values, which inherently involves uncertainty. They are represented in various ways, such as probability mass functions (pmf), cumulative distribution functions (cdf), samples, or probability quantiles (or intervals). 
Note that there might be other ways, such as named distributions (e.g.~Normal(1,2)) that are not currently supported by hubverse.

The \texttt{output\_type} and \texttt{output\_type\_id} columns in the hubverse model output format specify the forecast structure. In the \pkg{modelimportance} package, model outputs may only contain one \texttt{output\_type} of `mean', `median', `quantile', or `pmf': `mean' or `median' for point forecasts and `quantile' or `pmf' for probabilistic forecasts.
As aforementioned, \texttt{output\_type\_id} column identifies additional detailed information, such as specific quantile levels (e.g., ``0.1'', ``0.25'', ``0.5'', ``0.75'', and ``0.9'') for the `quantile' output type and categorical values (e.g., ``low'', ``moderate'', ``high'', and ``very high'') for the `pmf' output type. 
The predicted values for \texttt{pmf} are constrained to be between 0 and 1, indicating the probability at each categorical level, while they are unbounded numeric otherwise. Different output types correspond to different scoring rules for evaluating a model's prediction performance.
Table~\ref{tbl-pair-output-scoringrule} presents the output types and their associated scoring rules supported by the \pkg{modelimportance} package. 
We note that while the scoring rules are listed in the table using their conventional names, the package uses their negative values for evaluation (e.g., \(-\text{RSE}, -\text{WIS}\)) so that higher scores indicate better performance. This positively oriented scoring rule facilitates the interpretation of importance scores: positive values indicate that the model's inclusion in the ensemble improves ensemble performance, whereas negative values indicate that it worsens it. Further, using positive values to indicate improved ensemble performance aligns with the convention of Shapely values.

\begin{table}
\centering{
    \begin{tabular}{ll>{\raggedright\arraybackslash}p{10cm}}
    \toprule
    Output Type & Scoring Rule & Description\\
    \midrule
    mean & SE & Squared error (SE): the squared difference between the predicted value and the observed value\\
    median & AE & Absolute error (AE): the absolute difference between the predicted value and the observed value\\
    quantile & WIS & Weighted interval score (WIS): a quantile-based approximation of the continuous ranked probability score (CRPS) for evaluating quantile forecasts\\
    pmf & Log Score & Logarithm of the probability assigned to the true outcome (LogScore)\\
    \bottomrule
\end{tabular}
}
\caption{\label{tbl-pair-output-scoringrule}Pairs of output types and their associated scoring rules for evaluating prediction performance.}
\end{table}

\subsection{Oracle output data}\label{ch2subsec:oracle_output_data}

The \texttt{oracle\_output\_data} is a data frame that contains the ground truth values for the variables used to define modeling targets \citep{hubverse_docs, kerr2025coordinating}. 
Its name originates from the concept of an oracle making a perfect prediction formatted similarly to model output. This type of data must follow the oracle output format defined in the hubverse standard, which includes independent task ID columns (e.g., \texttt{location}, \texttt{target\_date}), the \texttt{output\_type} column specifying the output type of the predictions and an \texttt{oracle\_value} column for the observed values. As in the forecast data, if the \texttt{output\_type} is either \texttt{"quantile"} or \texttt{"pmf"}, the \texttt{output\_type\_id} column is often required to provide further identifying information.

The \texttt{model\_out\_tbl} and \texttt{oracle\_output\_data} must have the same task ID columns and \texttt{output\_type}, including \texttt{output\_type\_id} if necessary, as the \texttt{oracle\_output\_data} is joined with the \texttt{model\_out\_tbl} based on these fields in order to score the forecast performance.

\section{Method description and algorithms}\label{ch2sec:algorithms}

\begin{figure}[t]
\centering{
    \includegraphics[width=1\linewidth,height=\textheight,keepaspectratio]{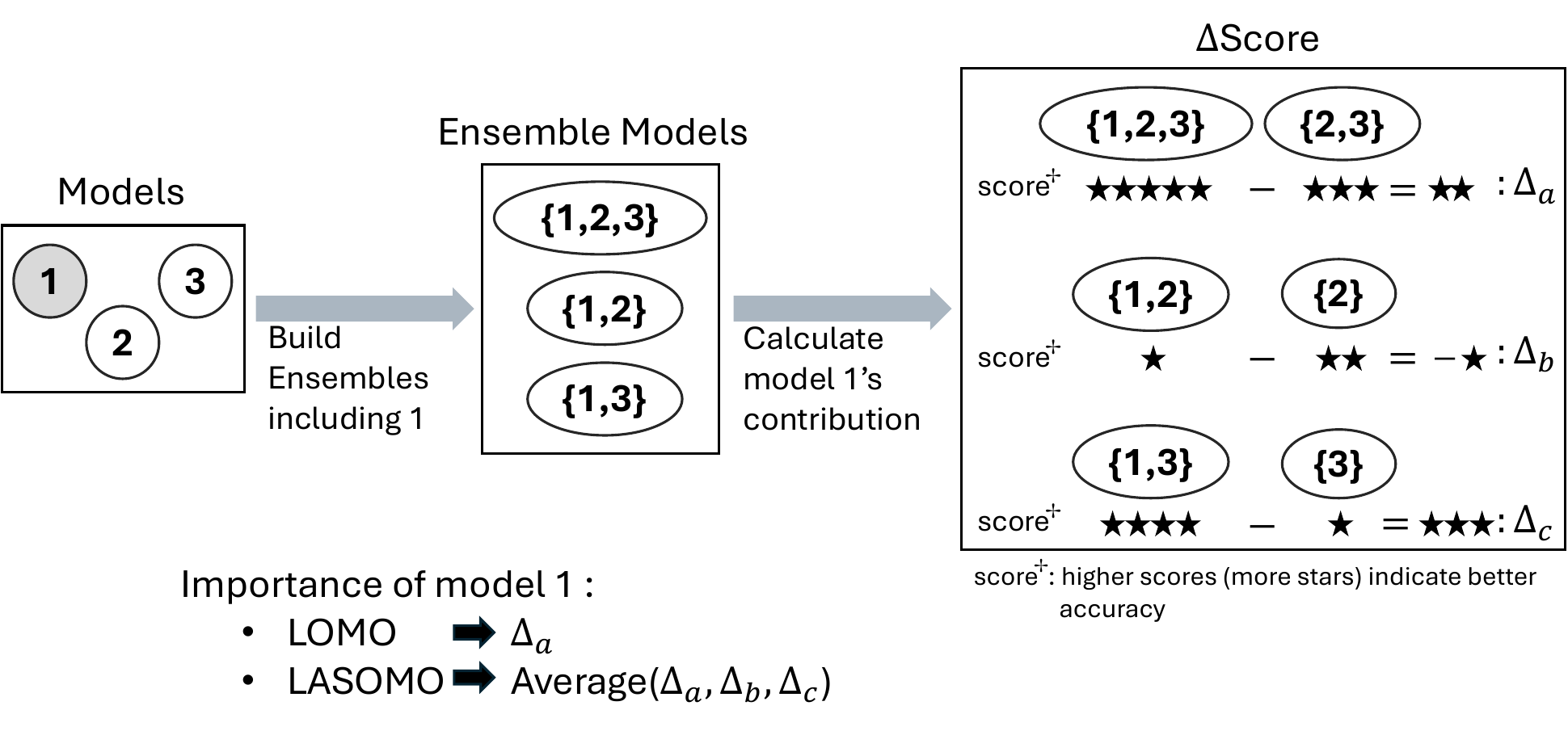}
}
\caption{\label{fig-concept}Conceptual illustration of measuring model
importance in a three-model setting. Each circle represents a
forecasting model with component model(s) shown inside. Stars indicate
the performance score measured by a positively oriented scoring rule
(e.g., \(-\)WIS), with more stars indicating better accuracy. LOMO
computes a single performance difference by removing the target model 1
from the full ensemble. LASOMO leverages multiple performance
differences across all subsets containing the model 1 and aggregates
them.}
\end{figure}

This section provides a brief description of the leave one model out
(LOMO) and leave all subsets of models out (LASOMO) algorithms, which
are used to compute the model importance score. The basic idea of
measuring the importance of each component model is to evaluate the
change in ensemble performance when that model is included or excluded
in the ensemble construction. More specifically, we compare the
performance of an ensemble with and without a specific model for a
specific task, and consider the difference in performance as the
importance of that model for that task (Figure~\ref{fig-concept}). We
apply this idea across many tasks and then average the task-level values
to obtain model-level summaries. The full theoretical development and
additional experiments are provided in \cite{kim2026beyond}.

\begin{algorithm}[b!]
\caption{Importance score calculation for one prediction task using leave one model out (LOMO) algorithm} 
\label{alg:lomo}
 \begin{spacing}{1.2}
  \begin{algorithmic}[1]
   \REQUIRE{
    \Statex $\bullet$ Set of $n$ individual models, ${\mathcal A}=\{1,2,\dots, n\},$ and their forecasts $\{F^1,F^2,...,F^n\}$
    \Statex $\bullet$ Truth value, $y$
    \Statex $\bullet$ Function $g$ to build an ensemble
    }
   \ENSURE Importance scores of all models, $\phi^{i,\text{lomo}} (i=1,2,\dots, n)$
    \STATE{Assign a positively oriented scoring rule $\psi$ to evaluate forecast skills, depending on the output type (\hyperref[tbl-pair-output-scoringrule]{Table~\ref{tbl-pair-output-scoringrule}})}
    \STATE{Create an ensemble forecast $F^{\mathcal A}$ using $g$: $F^{\mathcal A} \gets g(\{F^1,F^2,...,F^n\})$}
    \STATE{Evaluate $F^{\mathcal A}$ using $\psi$: $\psi(F^{\mathcal A},y).$}
    \FOR{each $i, (i = 1,2,..., n)$}
      \STATE {$F^{{\mathcal A}^{-i}} \gets g(\{F^j|\, j\ne i, j\in {\mathcal A}\})$}
      \STATE{Compute $\psi(F^{{\mathcal A}^{-i}},y).$}
      \STATE $\phi^{i,\text{lomo}} \gets \psi(F^{\mathcal A},y) - \psi(F^{{\mathcal A}^{-i}},y)$
    \ENDFOR
  \end{algorithmic}
 \end{spacing}
\end{algorithm}

LOMO involves creating an ensemble by excluding one component model from
the entire set of models. Let \({\mathcal A}\) be a set of \(n\) models
and \(F^i\) be a forecast produced by model \(i\), where
\(i = 1,2, \dots, n.\) Each ensemble excludes exactly one model while
including all the others. \(F^{{\mathcal A}^{-i}}\) denotes the ensemble
forecast constructed using all forecasts \(F^{\mathcal A}\) except
\(F^i\). Model \(i\)'s importance score using LOMO is calculated as the
difference in accuracy, as measured by a specific scoring rule, between
\(F^{{\mathcal A}^{-i}}\) and \(F^{\mathcal A}\)
(\hyperref[alg:lomo]{Algorithm~\ref{alg:lomo}}). For example, when
evaluating model 1 within an ensemble of three models (\(n=3\)), LOMO
creates an ensemble forecast \(F^{\{2,3\}}\) using only \(F^2\) and
\(F^3\). The performance of this reduced ensemble is then compared to
the full ensemble forecast \(F^{\{1,2,3\}}\), which incorporates all
three models. We note that a model can make an ensemble better or worse,
and thus the importance score for model 1 can be positive or negative
accordingly.

\begin{algorithm}[t!]
\caption{Importance score calculation for one prediction task using leave all subsets of models out (LASOMO) algorithm} 
\label{alg:lasomo}
 \begin{spacing}{1.2}
  \begin{algorithmic}[1]
   \REQUIRE{
    \Statex $\bullet$ Set of $n$ individual models, ${\mathcal A}=\{1,2,\dots, n\},$ and their forecasts $\{F^1,F^2,...,F^n\}$
    \Statex $\bullet$ Truth value, $y$
    \Statex $\bullet$ Function $g$ to build an ensemble
    \Statex $\bullet$ Subset weighting option, either `equal' or `perm\_based' }
   \ENSURE Importance scores of all models, $\phi^{i,\text{lasomo}} (i=1,2,\dots, n)$
   \STATE{Assign a positively oriented scoring rule $\psi$ to evaluate forecast skills, depending on the output type (\hyperref[tbl-pair-output-scoringrule]{Table~\ref{tbl-pair-output-scoringrule}})}
   \FOR{$i = 1 \text{ to } n$}
     \STATE $\phi^{i,\text{lasomo}} \gets 0$
     \STATE Make a list of non-empty subsets of $\mathcal A$ that does not contain $i$: $S_1, S_2, \cdots, S_{2^{n-1}-1}$
     \FOR{$j = 1 \text{ to } 2^{n-1}-1$}
       \STATE{Assign a weight to the subset $S_j$: }
         \IF{`equal' subset weighting is selected}
           \STATE $\gamma_{S_j} \gets \displaystyle\frac{1}{2^{n-1}-1}$
         \ELSE{ (\textit{`perm\_based' subset weighting})}
           \STATE {$\gamma_{S_j} \gets \displaystyle\frac{1}{(n-1)\binom{n-1}{|S_j|}}$}
         \ENDIF
       \STATE{$F^{S_j}\gets g(\{F^j|\, j\in S_j\})$}
       \STATE{$F^{{S_j}\cup \{i\}}\gets g(\{F^i, F^j|\, j\in S_j\})$}
       \STATE{Compute $\psi(F^{S_j},y)$ and
       $\psi(F^{{S_j}\cup\{i\}},y)$}
       \STATE {$\phi^{i,\text{lasomo}} \gets \phi^{i,\text{lasomo}} + \gamma_{S_j}\times\Big[ \psi(F^{{S_j}\cup\{i\}},y) -\psi(F^{S_j},y) \Big]$}
     \ENDFOR
   \ENDFOR
  \end{algorithmic}
 \end{spacing}
\end{algorithm}

On the other hand, LASOMO involves ensemble constructions from all
possible subsets of models. For each subset \(S\) that does not contain
the model \(i\), \(S \cup \{i\}\) plays the role of \({\mathcal A}\) in
the LOMO; the score associated with the subset \(S\) is the difference
of measures between \(F^S\) and \(F^{S \cup \{i\}}\). Then, all scores
are aggregated across all possible subsets that the model \(i\) does not
belong to (\hyperref[alg:lasomo]{Algorithm~\ref{alg:lasomo}}). For
example, using the earlier setup of three forecast models, LASOMO
considers three subsets, which we denote by \(S_1=\{2\}\),
\(S_2=\{3\}\), and \(S_3=\{2, 3\}\), to calculate the importance score
of model 1 (excluding all subsets that include model 1). The ensemble
forecasts \(F^{\{2\}}, F^{\{3\}}\), and \(F^{\{2,3\}}\) are then
compared to \(F^{\{1,2\}}, F^{\{1,3\}}\), and \(F^{\{1,2,3\}}\),
respectively. The performance differences attributable to model 1's
inclusion are aggregated, which results in the importance score for
model 1. We note that the subsets (e.g., \(S_1, S_2,\) and \(S_3\)) may
have different weights during the aggregation process.

The \pkg{modelimportance} package offers two weighting options for
subsets: one assigns equal (uniform) weights to all subsets, and the
other assigns weights based on their size, similar to the definition of
Shapley values, where a player's average contribution is aggregated over
all possible coalitions (or, equivalently, over all permutations of
players) (\cite{Shapley1953}). Users can choose one to evaluate the
contribution of each model in a manner suited to their preferred
framework. Uniform weighting may be preferred when each subset size is
of equal analytical interest, regardless of how many such subsets there
are, while size-based weighting may be preferred when users want to
maintain the original Shapley value interpretation, preventing numerous
medium-sized subsets from dominating the importance scores. A detailed
discussion of the differences between the two weighting schemes follows
in the next section.

\subsection{Comparison of weighting schemes in LASOMO}\label{ch2sec:comparison-weighting-schemes}

\begin{figure}[t]
\centering{
    \includegraphics[keepaspectratio]{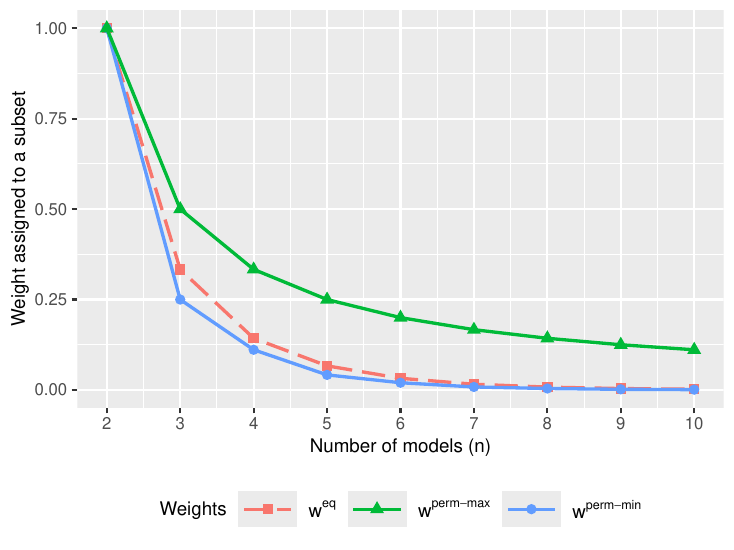}
}
\caption{\label{fig-lasomo-weights}Comparison of weights assigned to a subset. The plot shows the weights assigned to a subset as the number of models \(n\) increases from 2 to 10. The red dashed line represents the weights under the equal weighting scheme, while the blue and green lines represent the minimum and maximum weights, respectively, under the permutation-based weighting scheme. The minimum weight occurs when the subset size is around \((n-1)/2\), and the maximum weight occurs when the subset size is \(n-1\). As the number of models increases, the weights assigned by the two schemes become increasingly similar for mid-sized subsets whereas substantial differences remain for extreme-sized subsets.}
\end{figure}%

The differences in how the two weighting schemes influence the importance scores become more pronounced as the number of models increases. As described in \hyperref[alg:lasomo]{Algorithm~\ref{alg:lasomo}}, the formulas for their subset weights are \[\frac{1}{2^{n-1}-1} \quad\text{and}\quad \frac{1}{(n-1)\binom{n-1}{k}},\] where \(k\) is the size of each subset and \(n\) is the total number of models. The equal scheme (left formula) treats all subsets equally, so medium-sized subsets have considerable influence in the final result, as there are many such subsets. In contrast, the permutation-based scheme (right formula) adjusts the weights according to the subset size, giving the greatest weight to both the smallest and largest subsets while assigning small weights to the mid-sized subsets. Moreover, the weights assigned to the mid-sized subsets under the permutation-based approach decrease much faster with \(n\) than those under the equal weighting scheme (see \hyperref[ch2sec:appendix]{Appendix~\ref{ch2sec:appendix}} for details). Consequently, when \(n\) is large, middle-sized subsets play a dominant role in determining the importance scores under the equal weighting scheme, whereas extreme-sized subsets primarily drive the scores under the permutation-based weighting approach.

Overall, the difference between the two weighting schemes is likely to arise mainly from the the extreme-sized subsets when \(n\) is large.
This is because the weights given to the mid-sized subsets become increasingly similar, which are very small values on the order of \(10^{-3}\) even when \(n=8\), while the weights assigned to the smallest and largest subsets remain substantially different (Figure~\ref{fig-lasomo-weights}).

\section{Evaluating models with the model\_importance() function}\label{ch2sec:main-function}

In this section, we describe the usage of the function
\texttt{model\_importance()}, where multiple options are available to
customize the evaluation framework (Table~\ref{tbl-arguments1}).

\begin{table}
\centering{
\centering
\begingroup\fontsize{10.5}{12.5}\selectfont
    \begin{tabular}{l>{\raggedright\arraybackslash}p{3.8cm}>{\raggedright\arraybackslash}p{3.3cm}>{\raggedright\arraybackslash}p{2.8cm}}
    \toprule
    Argument & Description & Possible Values & Default\\
    \midrule
    \cellcolor{gray!10}{\texttt{forecast\_data}} & \cellcolor{gray!10}{Forecasts} & \cellcolor{gray!10}{Table of model output} & \cellcolor{gray!10}{N/A}\\
    \texttt{oracle\_output\_data} & Ground truth data & Table of oracle output & N/A\\
    \cellcolor{gray!10}{\texttt{ensemble\_fun}} & \cellcolor{gray!10}{Ensemble method} & \cellcolor{gray!10}{\texttt{"simple\_ensemble"}, \texttt{"linear\_pool"}} & \cellcolor{gray!10}{\texttt{"simple\_ensemble"}}\\
    \texttt{importance\_algorithm} & Algorithm to calculate importance & \texttt{"lomo", "lasomo"} & \texttt{"lomo"}\\
    \cellcolor{gray!10}{\texttt{subset\_wt}} & \cellcolor{gray!10}{Method for assigning weight to subsets when using LASOMO algorithm} & \cellcolor{gray!10}{\texttt{"equal", "perm\_based"}} & \cellcolor{gray!10}{\texttt{"equal"}}\\
    \addlinespace
    \texttt{min\_log\_score} & Minimum value to replace for log score & Non-positive numeric & -10\\
    \cellcolor{gray!10}{\texttt{...}} & \cellcolor{gray!10}{Optional arguments for \texttt{"simple\_ensemble"}} & \cellcolor{gray!10}{Varies} & \cellcolor{gray!10}{\texttt{agg\_fun="mean"}}\\
    \bottomrule
    \end{tabular}
\endgroup{}
}
\caption{\label{tbl-arguments1}Description of the arguments for the
\texttt{model\_importance()} function, including their purpose, possible
values, and default settings.}
\end{table}%

The \texttt{model\_importance()} function calculates the importance
scores of ensemble component models based on their contributions to
improving ensemble prediction accuracy for each prediction task. The
function requires a minimum of two models per task, as with only one
model, the importance score is not defined. The output of the function
is a single data frame of importance scores combined across all tasks.
If a model missed predictions for a specific task, an \texttt{NA} value
will be assigned for that task.

\begin{verbatim}
> model_importance(
        forecast_data, oracle_output_data, ensemble_fun, 
        importance_algorithm, subset_wt, min_log_score, ...
)
\end{verbatim}

The \texttt{forecast\_data} parameter is a data frame of predictions
that is automatically coerced to a \texttt{model\_out\_tbl} format (if
it is not one already), which is the standard S3 class model output
format defined by the hubverse convention. If coercion fails, users may
need to manually convert their data using the
\texttt{as\_model\_out\_tbl()} function from \pkg{hubUtils}.

The \texttt{oracle\_output\_data} is a data frame containing the actual
observed values of the variables used to specify modeling targets.
Further details on formatting and purpose are provided in
\hyperref[subsec:oracle_output_data]{Section~\ref{ch2subsec:oracle_output_data}}.

The \texttt{ensemble\_fun} argument specifies the ensemble method to use
when evaluating model importance, which relies on implementations in the
\pkg{hubEnsembles} package (\cite{Rpackage-hubEnsembles}). The
currently supported methods are \texttt{"simple\_ensemble"} and
\texttt{"linear\_pool"}. The \texttt{"simple\_ensemble"} method returns
the average of the predicted values from all component models per
prediction task defined by task IDs, \texttt{output\_type}, and
\texttt{output\_type\_id} columns. The default aggregation function for
this method is \texttt{"mean"}, but it can be customized by specifying
additional arguments through \texttt{...}, such as
\texttt{agg\_fun="median"}. When \texttt{"linear\_pool"} is selected,
the ensemble is a linear opinion pool of its component models. This
method supports only an \texttt{output\_type} of \texttt{"mean"},
\texttt{"quantile"}, or \texttt{"pmf"}.

The \texttt{importance\_algorithm} argument specifies the algorithm for
model importance calculation, which can be either \texttt{"lomo"} (leave
one model out) or \texttt{"lasomo"} (leave all subsets of models out).
The \texttt{subset\_wt} argument is employed only for the
\texttt{"lasomo"} algorithm. This argument has two options:
\texttt{"equal"} assigns equal weight to all subsets and
\texttt{"perm\_based"} assigns weight averaged over all possible
permutations, as in the formula of Shapley values
(\hyperref[alg:lasomo]{Algorithm~\ref{alg:lasomo}}). The default values
of \texttt{importance\_algorithm} and \texttt{subset\_wt} are
\texttt{"lomo"} and \texttt{"equal"}, respectively.

The \texttt{min\_log\_score} argument is relevant only for the
\texttt{output\_type} of \texttt{"pmf"}, which uses Log Score as a
scoring rule. It sets a minimum threshold for log scores to avoid issues
with extremely low probabilities assigned to the true outcome, which can
lead to undefined or negative infinite log scores. Any probability lower
than this threshold will be adjusted to this minimum value before
calculating the importance metric based on the log score. The default
value is set to -10, following the CDC FluSight thresholding convention
\cite{brooks2018nonmechanistic, reich_accuracy_2019}. Users may choose
a different value based on their practical needs.

The \texttt{model\_importance()} function returns an object of S3 class
\texttt{model\_imp\_tbl} of model important scores, with columns
\texttt{model\_id}, \texttt{reference\_date}, \texttt{output\_type}, and
\texttt{importance}, along with any task ID columns (e.g.,
\texttt{location}, \texttt{horizon}, and \texttt{target\_end\_date})
present in the input \texttt{forecast\_data}. Regardless of the original
column name for the forecast generation date in the
\texttt{forecast\_data}, it is standardized to \texttt{reference\_date}
in the output. This standardization enables consistent handling of the
forecast generation date across datasets, simplifying downstream
processing. The \texttt{importance} column contains the calculated
importance scores for each model and specific task, which are derived
from the specified algorithm and ensemble method.

\section{S3 class and methods}\label{ch2sec:s3-infrastructure}

We have defined a custom S3 class, \texttt{model\_imp\_tbl} that
represents the output of the \texttt{model\_importance()} function,
which extends the base \texttt{data.frame} class. Objects of this class
enable dispatch for class-specific methods for printing, summarizing,
and aggregating them across tasks.

\subsection{Print method}\label{print-method}

\texttt{print()} provides all computation results in a clear and
organized manner. It displays \texttt{model\_id} and \texttt{importance}
by grouped task IDs (e.g., \texttt{reference\_date}, \texttt{location},
and \texttt{target\_end\_date}) to facilitate easy interpretation of the
importance scores for each model across different tasks.

\subsection{Summary method}\label{summary-method}

\texttt{summary()} provides a concise summary of the importance scores,
including key statistics such as the number of models and tasks
evaluated. It also displays top-scoring models for a subset of tasks.
Additional summary details are available by specifying individual
elements of the summary object as follows:

\begin{itemize}
\item
  \texttt{.\$all\_tasks} lists all the unique tasks evaluated, which are
  defined by the combinations of task ID columns present in the input
  \texttt{forecast\_data}. This allows users to identify the scope of
  the evaluation.
\item
  \texttt{.\$model\_summary} provides each model's performance summary
  across tasks (e.g., the number of tasks each model submitted its
  forecast for, and the range of scores it achieved). This helps users
  understand the consistency and variability of each model's importance
  across different tasks.
\item
  \texttt{.\$task\_winners} identifies which model is the best for each
  task based on the highest importance score across the full set of
  tasks. From this information, users can quickly identify the most
  important model in each task.
\end{itemize}

\subsection{Aggregate method}\label{aggregate-method}

\texttt{aggregate()} allows users to obtain an overall importance score
for each model by aggregating its importance scores across all evaluated
tasks. This \texttt{model\_imp\_tbl} class-specific method provides
arguments that users can specify to customize the aggregation process,
including the handling of \texttt{NA} values and the choice of summary
function (Table~\ref{tbl-arguments2}).

\begin{table}[t]
\centering{
\centering
\begingroup\fontsize{10.5}{12.5}\selectfont
    \begin{tabular}{l>{\raggedright\arraybackslash}p{4cm}>{\raggedright\arraybackslash}p{3.7cm}>{\raggedright\arraybackslash}p{2cm}}
    \toprule
    Argument & Description & Possible Values & Default\\
    \midrule
    \cellcolor{gray!10}{\texttt{importance\_scores}} & \cellcolor{gray!10}{Model importance scores produced by \texttt{model\_importance()}} & \cellcolor{gray!10}{data frame} & \cellcolor{gray!10}{N/A}\\
    \texttt{by} & Grouping variable(s) for summarization & grouping variable(s) & \texttt{"model\_id"}\\
    \cellcolor{gray!10}{\texttt{na\_action}} & \cellcolor{gray!10}{Method to handle \texttt{NA} values} & \cellcolor{gray!10}{\texttt{"drop", "worst", "average"}} & \cellcolor{gray!10}{\texttt{"drop"}}\\
    \texttt{fun} & Function to summarize importance scores & summary function & \texttt{mean}\\
    \cellcolor{gray!10}{\texttt{...}} & \cellcolor{gray!10}{Optional arguments for \texttt{"fun"}} & \cellcolor{gray!10}{depends on \texttt{fun}} & \cellcolor{gray!10}{N/A}\\
    \bottomrule
    \end{tabular}
\endgroup{}
}
\caption{\label{tbl-arguments2}Description of the arguments for the
\texttt{aggregate()} function for \texttt{model\_imp\_tbl} objects,
including their purpose, possible values, and default settings.}
\end{table}%

The \texttt{by} argument specifies the grouping variable(s) for
summarization. Its default is \texttt{"model\_id"}, in which the
\texttt{aggregate()} method summarizes importance scores for each model.
Valid values for \texttt{by} include any combination of columns present
in the \texttt{importance\_scores} data frame.

The \texttt{na\_action} argument allows for specifying how to handle
\texttt{NA} values generated during importance score calculation for
each task; these values occur when a model did not contribute to the
ensemble prediction for a given task by missing its forecast submission.
Three options are available: \texttt{"worst"}, \texttt{"average"}, and
\texttt{"drop"}. In each specific prediction task, if a model has any
missing predictions, the \texttt{"worst"} option replaces the
\texttt{NA} values with the smallest value among other models'
importance metrics, while the \texttt{"average"} option replaces them
with the average of the other models' importance metrics in that task.
The \texttt{"drop"} option removes the \texttt{NA} values, which results
in the exclusion of the model from the evaluation for that task.

The \texttt{fun} argument specifies a function used to summarize
importance scores. An arithmetic mean (\texttt{fun\ =\ mean}) is the
default, but other summary functions (e.g., \texttt{fun\ =\ median}) or
user-defined functions may also be used. Additional arguments for the
summary function \texttt{fun} can also be passed via \texttt{...} if
needed (e.g., \texttt{fun\ =\ quantile,\ probs\ =\ 0.25} for a quartile
summary).

The output returned by \texttt{aggregate()} is a data frame with columns
\texttt{model\_id} and \\
\texttt{importance\_score\_\textless{}fun\textgreater{}}, where
\texttt{\textless{}fun\textgreater{}} is the name of the summary
function used (e.g.,\\
\texttt{importance\_score\_mean} when
\texttt{fun\ =\ mean}). The output is sorted in descending order of the
summarized importance scores,
\texttt{importance\_score\_\textless{}fun\textgreater{}}.

\section{Examples}\label{ch2sec:examples}

In this section, we illustrate how to use the
\texttt{model\_importance()} function to evaluate the importance of
component models within an ensemble. The examples show various
combinations of the arguments described in
\hyperref[sec:main-function]{Section~\ref{ch2sec:main-function}}. Example
forecast and target data are originally sourced from the
\pkg{hubExamples} package, which provides sample datasets for multiple
modeling hubs in the hubverse format.

\subsection{Example data}\label{ch2sec:example-data}
Our example forecast data contains short-term predictions of weekly incident influenza hospitalizations in the US for Massachusetts (FIPS code 25) and Texas (FIPS code 48), generated on November 19, 2022. These forecasts are made for two target end dates, November 26, 2022 (horizon 1), and December 10, 2022 (horizon 3), and were produced by three
models: `Flusight-baseline', `MOBS-GLEAM\_FLUH', and `PSI-DICE'. The output type is \texttt{median} and the \texttt{output\_type\_id} column
has \texttt{NA}s as no further specification is required for this output type. 
We have modified the example data slightly by removing some
forecasts to demonstrate the handling of missing values. 
Therefore, MOBS-GLEAM\_FLUH's forecast for Massachusetts on November 26, 2022, and PSI-DICE's forecast for Texas on December 10, 2022, are missing. 
We emphasize that this modification is made randomly and artificially to illustrate the impact of different handling approaches for missing forecasts and is not intended to provide a formal evaluation of model quality or performance.

\vspace{0.35cm}
\begin{verbatim}
> forecast_data |>
        knitr::kable(format = "latex", booktabs = TRUE) |>
        kable_styling(latex_options = c("scale_down", "hold_position"))
\end{verbatim}

\begin{table}[H]
\centering
\resizebox{\ifdim\width>\linewidth\linewidth\else\width\fi}{!}{
    \begin{tabular}{lllrllllr}
    \toprule
    model\_id & reference\_date & target & horizon & location & target\_end\_date & output\_type & output\_type\_id & value\\
    \midrule
    Flusight-baseline & 2022-11-19 & wk inc flu hosp & 1 & 25 & 2022-11-26 & median & NA & 51\\
    Flusight-baseline & 2022-11-19 & wk inc flu hosp & 3 & 25 & 2022-12-10 & median & NA & 51\\
    Flusight-baseline & 2022-11-19 & wk inc flu hosp & 1 & 48 & 2022-11-26 & median & NA & 1052\\
    Flusight-baseline & 2022-11-19 & wk inc flu hosp & 3 & 48 & 2022-12-10 & median & NA & 1052\\
    MOBS-GLEAM\_FLUH & 2022-11-19 & wk inc flu hosp & 3 & 25 & 2022-12-10 & median & NA & 43\\
    MOBS-GLEAM\_FLUH & 2022-11-19 & wk inc flu hosp & 1 & 48 & 2022-11-26 & median & NA & 1072\\
    MOBS-GLEAM\_FLUH & 2022-11-19 & wk inc flu hosp & 3 & 48 & 2022-12-10 & median & NA & 688\\
    PSI-DICE & 2022-11-19 & wk inc flu hosp & 1 & 25 & 2022-11-26 & median & NA & 90\\
    PSI-DICE & 2022-11-19 & wk inc flu hosp & 3 & 25 & 2022-12-10 & median & NA & 159\\
    PSI-DICE & 2022-11-19 & wk inc flu hosp & 1 & 48 & 2022-11-26 & median & NA & 1226\\
    \bottomrule
    \end{tabular}}
\end{table}

The \texttt{forecast\_data} output was rendered using the \pkg{knitr} \citep{Rpackage-knitr} package and styled with \pkg{kableExtra} \citep{Rpackage-kableExtra} to improve readability, including automatic scaling to fit within page width.

The corresponding target data contains the observed hospitalization counts for these dates and locations.

\begin{verbatim}
> target_data
\end{verbatim}

\begin{verbatim}
# A tibble: 4 x 4
  target_end_date target          location oracle_value
  <date>          <chr>           <chr>           <dbl>
1 2022-11-26      wk inc flu hosp 25                221
2 2022-11-26      wk inc flu hosp 48               1929
3 2022-12-10      wk inc flu hosp 25                578
4 2022-12-10      wk inc flu hosp 48               1781
\end{verbatim}

\begin{figure}[t]
\centering{
    \includegraphics[width=0.8\textwidth]{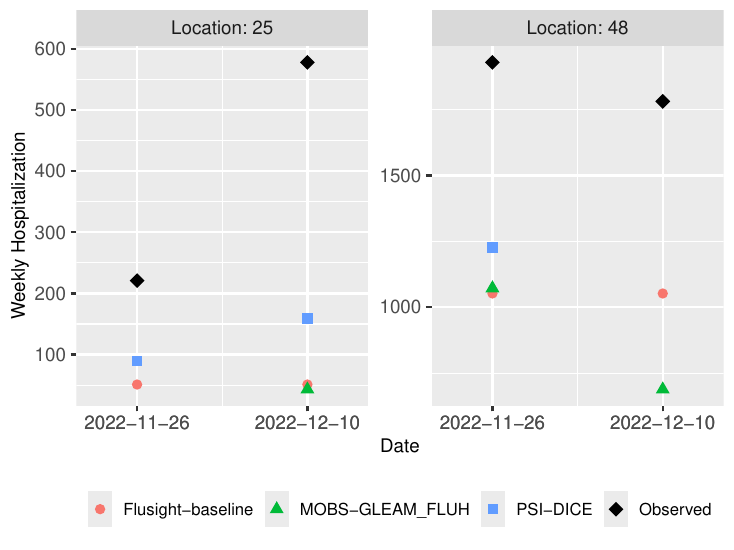}
}
\caption{\label{fig-example-median-lomo}Plot of three point forecasts (median) and the eventually observed values from the
\texttt{forecast\_data} and \texttt{target\_data} for weekly incident influenza hospitalizations in Massachusetts (FIPS code 25) and Texas (FIPS code 48). Colored dots indicate the forecasts by three models, generated on November 19, 2022. Open black circles indicate the eventually observed values. MOBS-GLEAM\_FLUH's forecast for Massachusetts on November 26, 2022, and PSI-DICE's forecast for Texas on December 10, 2022, are not shown. These forecasts were manually excluded from the original example dataset for demonstration purposes. }
\end{figure}

When comparing the ground truth data and model predictions, we can see
that forecasts for December 10, 2022 show larger deviations from the
observed values compared to those for November 26, 2022. Thus, as we
expect, prediction errors increase at longer horizons due to greater
uncertainty. Additionally, the forecasts for Massachusetts are
relatively more accurate compared to those for Texas, which tend to have
higher errors (Figure~\ref{fig-example-median-lomo}).

\subsection{Evaluation using LOMO algorithm}\label{ch2sec:example-lomo}

We quantify the contribution of each model within the ensemble using the
\texttt{model\_importance()} function. The following code evaluates the
importance of each ensemble member in the simple mean ensemble using the
LOMO algorithm.

\begin{verbatim}
> scores_lomo <- model_importance(
        forecast_data = forecast_data,
        oracle_output_data = target_data,
        ensemble_fun = "simple_ensemble",
        importance_algorithm = "lomo"
)
\end{verbatim}

This call also generates informative messages that summarize the input
data, including the number of dates on which forecasts were produced,
the number of models and their ids, and whether the prediction tasks
meet the minimum model requirement, alongside a note on how to use
parallel processing to speed up the computation when there are many
models and tasks. A print out of these messages is shown below.

\begin{verbatim}
Evaluating forecasts from 2022-11-19 to 2022-11-19  (a total of 1 forecast date(s)).

The available model IDs are:
     Flusight-baseline
     MOBS-GLEAM_FLUH
     PSI-DICE 
(a total of 3 models)

Note: This function uses 'furrr' and 'future' for parallelization.
To enable parallel execution, please set future::plan(multisession).

All tasks meet the minimum model requirement of 2 models.
\end{verbatim}

The function output is a data frame containing model ids and their
corresponding importance scores for each prediction task, along with
task id columns. (Note that we rounded the importance scores and renamed
the columns for better readability in the output below, but the original
column names are retained in the actual output data frame.)

\begin{verbatim}
> print(
        scores_lomo |> 
          mutate(importance = round(importance, 2)) |>
          rename(ref_date = reference_date, h=horizon, 
                 loc=location, t_end_date=target_end_date, 
                 o_type=output_type, imp=importance)
)
\end{verbatim}

\begin{verbatim}
Model importance result by task
---------------------------------
            model_id   ref_date          target h loc t_end_date o_type     imp
1  Flusight-baseline 2022-11-19 wk inc flu hosp 1  25 2022-11-26 median  -19.50
2    MOBS-GLEAM_FLUH 2022-11-19 wk inc flu hosp 1  25 2022-11-26 median      NA
3           PSI-DICE 2022-11-19 wk inc flu hosp 1  25 2022-11-26 median   19.50
4  Flusight-baseline 2022-11-19 wk inc flu hosp 1  48 2022-11-26 median  -32.33
5    MOBS-GLEAM_FLUH 2022-11-19 wk inc flu hosp 1  48 2022-11-26 median  -22.33
6           PSI-DICE 2022-11-19 wk inc flu hosp 1  48 2022-11-26 median   54.67
7  Flusight-baseline 2022-11-19 wk inc flu hosp 3  25 2022-12-10 median  -16.67
8    MOBS-GLEAM_FLUH 2022-11-19 wk inc flu hosp 3  25 2022-12-10 median  -20.67
9           PSI-DICE 2022-11-19 wk inc flu hosp 3  25 2022-12-10 median   37.33
10 Flusight-baseline 2022-11-19 wk inc flu hosp 3  48 2022-12-10 median  182.00
11   MOBS-GLEAM_FLUH 2022-11-19 wk inc flu hosp 3  48 2022-12-10 median -182.00
12          PSI-DICE 2022-11-19 wk inc flu hosp 3  48 2022-12-10 median      NA
\end{verbatim}

For models that missed forecasts for certain tasks, \texttt{NA} values
were assigned in the importance column for those tasks.

Calling \texttt{summary()} shows that three models were used and four
tasks were evaluated, along with a preview of the top-performing model
for each task.

\begin{verbatim}
> summary(scores_lomo)
\end{verbatim}

\begin{verbatim}
=== Summary of importance scores by task ===
Number of models: 3
Number of tasks: 4

=== Top scoring model by task for a subset of tasks ========================================
target horizon location target_end_date top_model importance
wk inc flu hosp 1 25 2022-11-26 PSI-DICE 19.50
wk inc flu hosp 1 48 2022-11-26 PSI-DICE 54.67
wk inc flu hosp 3 25 2022-12-10 PSI-DICE 37.33
--------------------------------------------
* More details are available in the summary object (e.g., $all_tasks, $model_summary,
$task_winners).
\end{verbatim}

As indicated in the output, more details about the summary are available
through the summary object's elements as follows.

\begin{verbatim}
> s <- summary(scores_lomo)
> s$all_tasks
\end{verbatim}

\begin{verbatim}
           target horizon location target_end_date
1 wk inc flu hosp       1       25      2022-11-26
2 wk inc flu hosp       1       48      2022-11-26
3 wk inc flu hosp       3       25      2022-12-10
4 wk inc flu hosp       3       48      2022-12-10
\end{verbatim}

Each row represents a unique combination of task IDs, from which we
verify that four different tasks were evaluated.

\begin{verbatim}
> s$model_summary
\end{verbatim}

\begin{verbatim}
           model_id n_tasks min_importance max_importance n_NA
1 Flusight-baseline       4         -32.33         182.00    0
2   MOBS-GLEAM_FLUH       4        -182.00         -20.67    1
3          PSI-DICE       4          19.50          54.67    1
\end{verbatim}

We observe that `Flusight-baseline' submitted forecasts for all four
tasks (n\_NA = 0), while `MOBS-GLEAM\_FLUH' and `PSI-DICE' submitted
forecasts for only three tasks due to one missing forecast (n\_NA = 1).
Each model's importance scores vary across tasks. `Flusight-baseline'
shows the largest range of scores that includes a negative minimum value
and a positive maximum value, while `MOBS-GLEAM\_FLUH' and `PSI-DICE'
have scores that are all positive or all negative across the three tasks
they submitted forecasts for.

\begin{verbatim}
> s$task_winners
\end{verbatim}

\begin{verbatim}
           target horizon location target_end_date         top_model max_score
1 wk inc flu hosp       1       25      2022-11-26          PSI-DICE     19.50
2 wk inc flu hosp       1       48      2022-11-26          PSI-DICE     54.67
3 wk inc flu hosp       3       25      2022-12-10          PSI-DICE     37.33
4 wk inc flu hosp       3       48      2022-12-10 Flusight-baseline    182.00
\end{verbatim}

Models with the highest importance scores for each task are identified
in the \texttt{top\_model} column with their importance score in the
\texttt{max\_score} column. `PSI-DICE' is the best model for three out
of the four tasks, while `Flusight-baseline' is the best for the
remaining task.

Visualization can be performed using functions from \pkg{ggplot2}. The
following example shows a bar plot of importance scores across models
and tasks, with panels faceted by combinations of task ID values.

\begin{verbatim}
> ggplot(scores_lomo, aes(x = model_id, y = importance, fill = model_id)) +
        geom_col() +
        coord_flip() +
        geom_hline(yintercept = 0, color = "black", linewidth = 0.25) +
        facet_grid(cols = vars(target, horizon, location, target_end_date), 
                   scales = "free_x") +
        labs(x = "Model ID", y = "Importance Score", 
             title = "Model Importance by Task") +
        scale_x_discrete(labels = function(x) gsub("[-_]", "-\n", x)) +
        theme(axis.text.x = element_text(angle = 90, hjust = 1, vjust = 0.5),
              panel.spacing.x = unit(0.5, "lines"), legend.position = "none")
\end{verbatim}
\vspace{0.1cm} 
\begin{center}
    \includegraphics[width=0.9\textwidth]{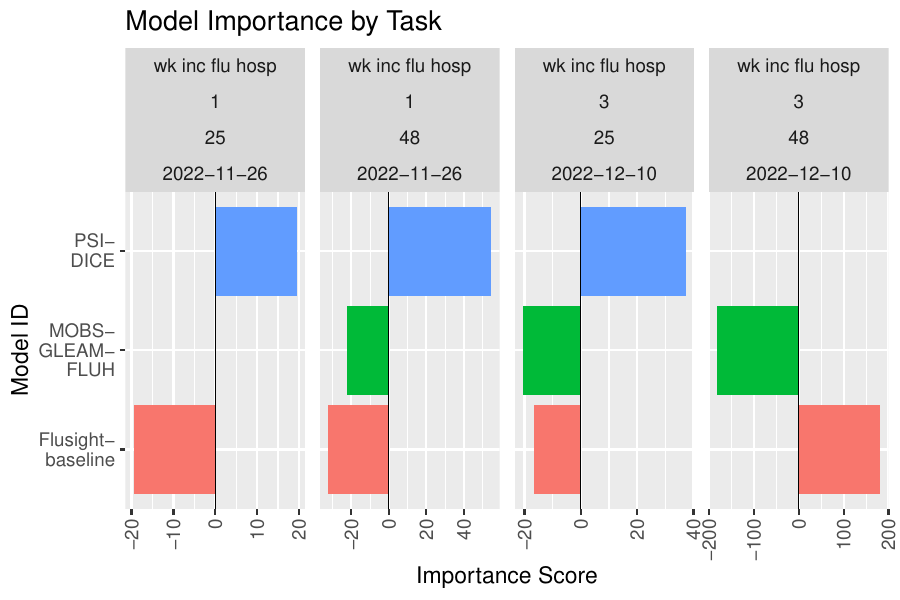}
\end{center}

\vspace{0.35cm} 

We aggregate the importance scores for each model by averaging across
all tasks. \texttt{NA} values are removed during the averaging process
by setting the \texttt{na\_action} argument to \texttt{"drop"}.

\begin{verbatim}
> aggregate(scores_lomo, by = "model_id", na_action = "drop", fun = mean) 
\end{verbatim}

\begin{verbatim}
Overall model importance across tasks
---------------------------------------- 
# A tibble: 3 x 2
  model_id          importance_score_mean
  <chr>                             <dbl>
1 PSI-DICE                           37.2
2 Flusight-baseline                  28.4
3 MOBS-GLEAM_FLUH                   -75  
\end{verbatim}

\vspace{0.35cm} 

The results show that, overall, the model `PSI-DICE' has the highest
importance score, followed by `Flusight-baseline' and
`MOBS-GLEAM\_FLUH'. That is, `PSI-DICE' contributes the most to
improving the ensemble's predictive performance, whereas
`MOBS-GLEAM\_FLUH', which has a negative score, detracts from the
ensemble's performance. The low importance score of `MOBS-GLEAM\_FLUH'
is mainly due to a substantially larger prediction error for Texas on
the target end date of December 10, 2022, compared to other models,
while its missing forecast for Massachusetts for November 26, 2022, was
not factored into the evaluation. This single large error significantly
affected its contribution score.

Another approach to handling missing values is to use the
\texttt{"worst"} option for \texttt{na\_action}, which replaces missing
values with the worst (i.e., minimum) score among the other models for
the same task.

\begin{verbatim}
> aggregate(scores_lomo, by = "model_id", na_action = "worst", fun = mean) 
\end{verbatim}

\begin{verbatim}
Overall model importance across tasks
---------------------------------------- 
# A tibble: 3 x 2
  model_id          importance_score_mean
  <chr>                             <dbl>
1 Flusight-baseline                  28.4
2 PSI-DICE                          -17.6
3 MOBS-GLEAM_FLUH                   -61.1
\end{verbatim}

\vspace{0.35cm} 

The results show that the importance score of `Flusight-baseline' is
unchanged because it has no missing forecasts. We observe that the
importance score of `PSI-DICE', which was previously positive, has now
decreased to a negative value when compared to the evaluation using the
\texttt{"drop"} option for \texttt{na\_action}. Moreover,
`MOBS-GLEAM\_FLUH' still ranks the lowest, but the importance score has
increased. This change is related to the varying forecast accuracy
across different tasks. For the target end date of November 26, 2022, in
Massachusetts, most forecasts are relatively accurate. Thus, even if the
`MOBS-GLEAM\_FLUH' is assigned the worst value of importance score for
its missing forecast, including this value in the averaging is not
detrimental to the overall importance metric; rather, it is more
beneficial than excluding it. In contrast, for the target end date of
December 10, 2022, in Texas, the forecasts have much larger errors
across the board, and assigning the worst value of importance score to
the missing forecast of `PSI-DICE' in this task has a detrimental effect
on averaging importance scores. This is because the scale of the
importance scores is influenced by the magnitude of the prediction
errors: for tasks with small errors, the scores remain moderate, while
tasks with large errors can yield importance scores of much greater
magnitude.

It is also possible to impute the missing scores with intermediate
values by assigning the average importance scores of other models in the
same task. This strategy may offer a more balanced trade-off by
mitigating the influence of the missing data without overly penalizing
or overlooking them.

\begin{verbatim}
> aggregate(scores_lomo, by = "model_id", na_action = "average", fun = mean) 
\end{verbatim}

\begin{verbatim}
Overall model importance across tasks
---------------------------------------- 
# A tibble: 3 x 2
  model_id          importance_score_mean
  <chr>                             <dbl>
1 Flusight-baseline                  28.4
2 PSI-DICE                           27.9
3 MOBS-GLEAM_FLUH                   -56.2
\end{verbatim}

\subsection{Evaluation using LASOMO algorithm}\label{ch2sec:example-lasomo}

We now demonstrate the use of the LASOMO algorithm for evaluating model
importance. Since we explored the difference of \texttt{na\_action}
options in the previous LOMO example
(\hyperref[sec:example-lomo]{Section~\ref{ch2sec:example-lomo}}), we focus
on options for \texttt{subset\_wt}, which specifies how weights are
assigned to subsets of models when calculating importance scores, with
\texttt{na\_action} fixed to \texttt{"drop"}.

The following code and corresponding outputs illustrate the evaluation
using each weighting scheme.

\begin{verbatim}
> scores_lasomo_eq <- model_importance(
        forecast_data = forecast_data,
        oracle_output_data = target_data,
        ensemble_fun = "simple_ensemble",
        importance_algorithm = "lasomo",
        subset_wt = "equal"
)
> aggregate(scores_lasomo_eq, by = "model_id", na_action = "drop", fun = mean)
\end{verbatim}

\begin{verbatim}
Overall model importance across tasks
---------------------------------------- 
# A tibble: 3 x 2
  model_id          importance_score_mean
  <chr>                             <dbl>
1 PSI-DICE                           47.4
2 Flusight-baseline                  24.3
3 MOBS-GLEAM_FLUH                   -79.8
\end{verbatim}

\vspace{0.35cm} 

\begin{verbatim}
> scores_lasomo_perm <- model_importance(
        forecast_data = forecast_data,
        oracle_output_data = target_data,
        ensemble_fun = "simple_ensemble",
        importance_algorithm = "lasomo",
        subset_wt = "perm_based"
)
> aggregate(scores_lasomo_perm, by = "model_id", na_action = "drop", fun = mean)
\end{verbatim}

\begin{verbatim}
Overall model importance across tasks
---------------------------------------- 
# A tibble: 3 x 2
  model_id          importance_score_mean
  <chr>                             <dbl>
1 PSI-DICE                           44.8
2 Flusight-baseline                  25.3
3 MOBS-GLEAM_FLUH                   -78.6
\end{verbatim}
\vspace{0.2cm}

In this example, there are only three models (\(n = 3\)), and the
weights do not differ significantly between the two weighting schemes.
Therefore, the resulting outputs show little difference. However, in
general, with a larger number of models, the two weighting schemes may
yield quite different importance scores for each model, as discussed in
\hyperref[sec:comparison-weighting-schemes]{Section~\ref{ch2sec:comparison-weighting-schemes}}.

\vspace{0.2cm}

In this section, we explored each component model's contribution to the ensemble accuracy with only three models. 
An extensive application in more complex scenarios with larger ensembles is presented in our companion methodological paper \citep{kim2026beyond}.

It should be noted that the example presented here is designed for illustration purposes to demonstrate the use of the proposed software, and the results should not be interpreted as an authoritative evaluation of model performance. A comprehensive analysis in \citep{kim2026beyond} shows that many models, including the `MOBS-GLEAM\_FLUH' model, are assessed as more important than the baseline model.

\section{Computational complexity}\label{ch2sec:computational-complexity}

This section describes the computational complexity of LOMO and LASOMO
algorithms implemented in the package, depending on the numbers of
ensemble component models and prediction tasks. We conducted a
computational experiment using simulated data with a point forecast
(`median'). The execution time was measured for both algorithms across
varying numbers of models and tasks. We also compared the execution
times between sequential and multisession computing environments to
demonstrate the efficiency of parallelization in handling
computationally intensive tasks.

We performed all experiments on a machine running macOS 15 with an Apple
M4 chip and 16 GB RAM under R version 4.4.3. Parallel computations were
implemented using the \pkg{future} (\texttt{multisession} backend) with
four worker processes. The execution times were recorded in seconds.

\begin{figure}[t]
\centering{
\includegraphics[width=1\linewidth,height=\textheight,keepaspectratio]{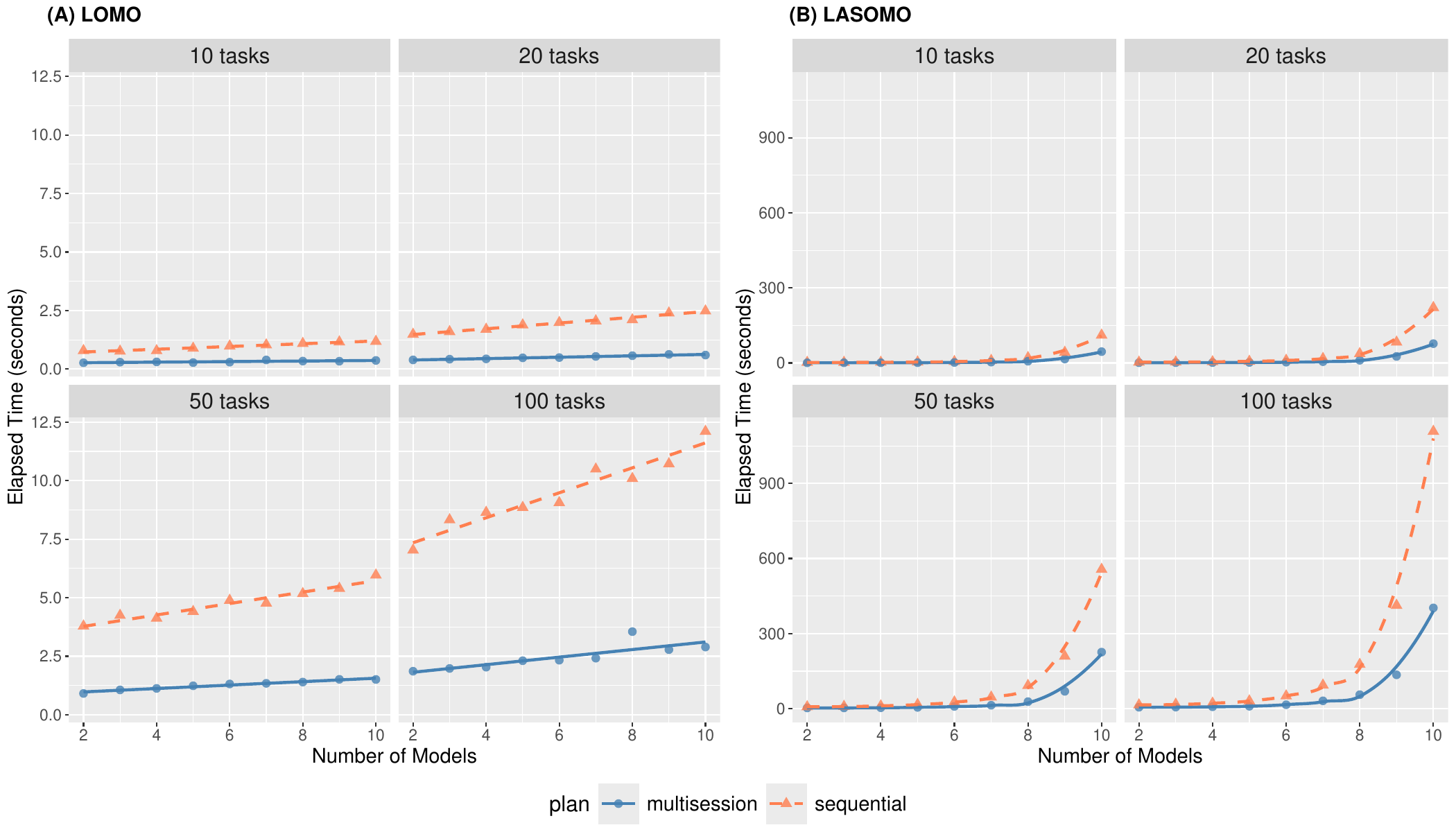}
}
\caption{\label{fig-feasibility-comparison}Runtime to compute model
importance scores for point predictions by number of ensemble component
models (ranging from 2 to 10) and number of prediction tasks (10, 20,
50, and 100) under sequential and parallel execution modes (multisession
backend, with four workers) for LOMO and LASOMO algorithms. Each point
represents the elapsed time (in seconds) for a given number of models,
while lines (solid/dashed) show the overall trend fitted to the
observations. The scale of the \(y\)-axis is different between the two
algorithms.}
\end{figure}%

Figure~\ref{fig-feasibility-comparison} illustrates the computational
runtime for four different numbers of prediction tasks (10, 20, 50, and
100) and varying numbers of models (ranging from 2 to 10) for both the
LOMO and LASOMO algorithms under sequential and parallel execution
modes. The scale of the \(y\)-axis approximately ranges from 0 to 12
seconds for LOMO and from 0 to 1110 seconds for LASOMO. Overall, the
time needed to compute model importance scores increases with the number
of models and tasks for both algorithms. Specifically, when the number
of tasks is 10 or 20, the difference in computation time between
sequential and parallel executions is not significant. However, as the
number of prediction tasks rises to 50, the difference becomes more
noticeable. The speedup from parallel execution relative to sequential
execution is approximately 4-fold for LOMO in these settings of 10, 20,
50, and 100 prediction tasks, which is consistent with the use of four
worker processes. Similarly, for LASOMO, a speedup of approximately
3-fold is observed across these settings, which is less than the 4-fold
speedup observed with LOMO due to the heavier computational intensity of
LASOMO. Nevertheless, the efficiency of parallel execution is more
pronounced for LASOMO than for LOMO, particularly when eight or more
models are involved in the ensemble construction and the number of
prediction tasks is large.

\begin{figure}[t]
\begin{minipage}{0.5\linewidth}
\subcaption{LOMO}\label{fig-feasibility-parallel-lomo}
\includegraphics[width=\linewidth]{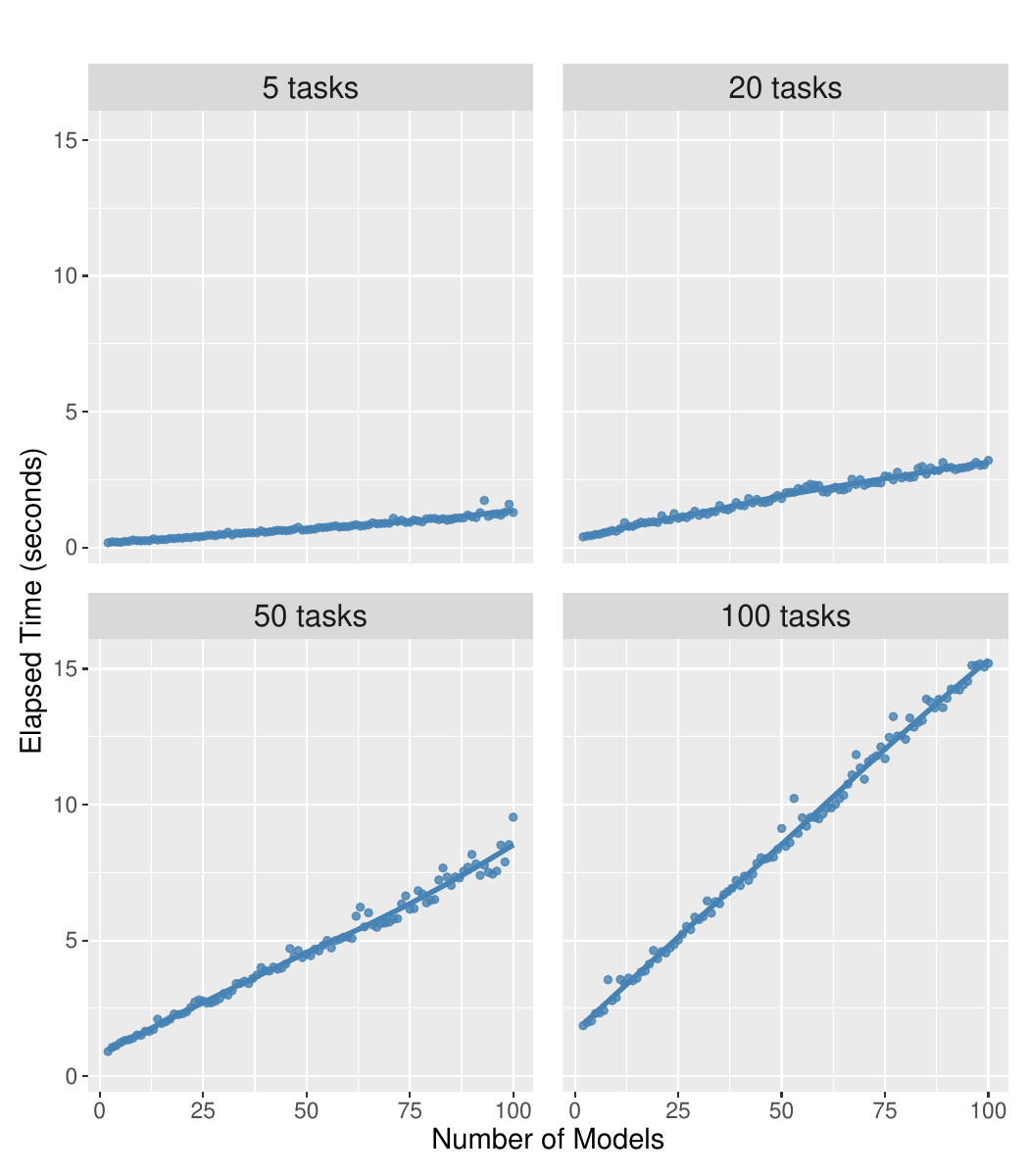}
\end{minipage}%
\hspace{0.01\linewidth}%
\begin{minipage}{0.5\linewidth}
\subcaption{LASOMO}\label{fig-feasibility-parallel-lasomo}
\includegraphics[width=\linewidth]{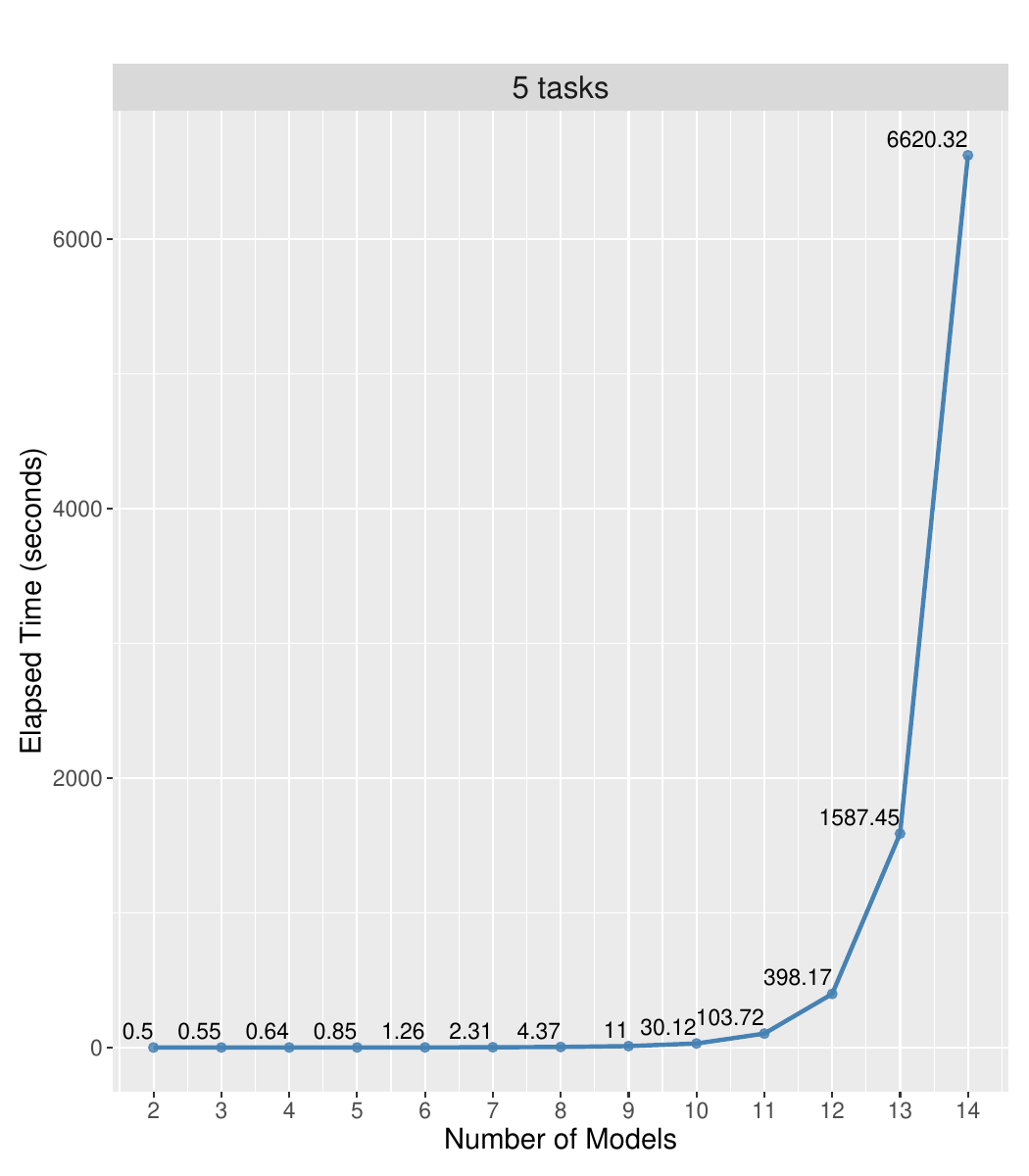}
\end{minipage}%
\caption{Runtime to compute model importance scores for point predictions by number of ensemble component models and prediction tasks during parallel execution via multisession backend. Each dot represents the elapsed time (in seconds) for a given number of models. For LOMO, the number of models evaluated varies from 2 to 100 for each 5, 20, 50, and 100 prediction tasks. The fitted lines illustrate the trend of the observations. For LASOMO, the number of models evaluated varies from 2 to 14 while fixing 5 prediction tasks due to exponentially growing computational intensity with the number of models. The exact elapsed time is shown for LASOMO, which captures the steep increase in runtime as the number of models grows, even before the high jump occurs with models over 12.}
\label{fig-feasibility-parallel}
\end{figure}

Theoretically, the computational complexity of LOMO and LASOMO
algorithms is \(O(t \cdot n)\) and \(O(t\cdot 2^n)\), respectively,
where \(n\) is the number of models and \(t\) is the number of
prediction tasks. However, the observations from our computational
experiment in the sequential execution mode exhibited slight
discrepancies from these theoretical complexities, under a limited
setting involving between 2 and 10 models and 10, 20, 50, and 100
prediction tasks. While the observed runtimes showed linear growth with
respect to the number of tasks for both algorithms, consistent with the
theoretical complexities, the growth with respect to the number of
models is sublinear for LOMO and slower than the expected exponential
for LASOMO. This deviation from the theoretical patterns is likely due
to the practical efficiency of the implementations, including shared
computation. Specifically, once the ensemble predictions are computed,
the importance scores are derived from these predictions without needing
to recompute the ensemble for each model. It should be noted that the
scaling increment and ratio between adjacent values of \(n\) in LOMO and
LASOMO, respectively, is gradually increasing, which implies the
empirical complexity is approaching the theoretical complexity with the
number of models. However, the execution time scales much more rapidly
with the number of models than the theoretical complexities, notably in
LASOMO, as larger sub-ensembles require more expensive evaluation and
scoring operations across all prediction tasks as well as larger
intermediate objects. That is, the theoretical complexity does not
factor in subset size, which does, in fact, impact the computational
speed in practice. This phenomenon is also observed even in the parallel
execution setting (Figure \ref{fig-feasibility-parallel-lasomo}).

We further explored computational feasibility, focusing on the parallel execution mode (Figure \ref{fig-feasibility-parallel}). We observe that, in LOMO, the elapsed time increases linearly with the number of models and the rate of increase depends on the number of tasks, with a steeper increase as the number of tasks grows. In contrast, in LASOMO, the elapsed time increases much more rapidly as the ensemble size grows. In particular, when the number of models exceeds 12, it takes more than 20 minutes to compute the importance scores, even with fixed 5 prediction tasks. It is because the number of subsets of models that need to be evaluated in LASOMO grows exponentially with the number of models, leading to a rapidly rising computational intensity. This observation highlights the trade-off between the comprehensiveness of the LASOMO algorithm and its computational challenges in practice, particularly with a large number of models.

\section{Implementation and availability}\label{ch2sec:implementation-and-availability}

The \pkg{modelimportance} package is implemented in \proglang{R} and currently distributed via GitHub under the MIT license (\url{https://github.com/mkim425/modelimportance}), with a CRAN release planned soon. 
We conducted unit tests using the \pkg{testthat} package \citep{Rpackage-testthat} to verify that all inputs and outputs are properly formatted and ensure that all functions work correctly as expected, including those used internally. 
We also performed continuous integration testing using GitHub Actions to maintain functionality across platforms, including Windows, macOS, and Linux. Integrated GitHub Action, we employed the \pkg{lintr} package \citep{Rpackage-lintr} to maintain code quality and detect potential issues, and the Air formatter \citep{air_formatter} to ensure consistent code style across the codebase. 
All code changes were systematically reviewed by fellow team members, and this enhanced reliability.

\section{Summary and discussion}\label{ch2sec:discussion}

Multi-model ensemble forecasts often provide better accuracy and robustness than single models, and are widely used in decision-making and policy planning across various domains. The contribution of each component model to the accuracy of the ensemble depends on its own unique characteristics. The \pkg{modelimportance} package enables the quantification of the value that each component model adds to the ensemble performance in different evaluation contexts.

In the example analysis, we showed the package workflow using a simple example dataset specifically designed solely to demonstrate our package's capabilities; thus the results should not be taken as a definitive assessment of the models. A more complete model importance analysis is provided in our companion methodological paper \citep{kim2026beyond}.

The primary \texttt{model\_importance()} function returns an S3 object of class \texttt{model\_imp\_tbl} of tabular data containing component models and their importance metrics. Users are given a choice of ensemble methods, model importance algorithms (either LOMO or LASOMO), and options to handle missing values. 
These features enable the package to serve as a versatile tool to aid collaborative efforts to construct an effective ensemble model across a wide range of forecasting tasks. 
We note that unit testing with continuous integration ensures the reliability of all functions and the overall quality of code across multiple platforms.

The \pkg{modelimportance} package still has several areas in which it
may be enhanced. Namely, although the package currently supports four
different output types (`mean', `median', `quantile, and 'pmf'), other
output types are widely used in practice. For example, the `sample'
output type is commonly used in the US Flu Scenario Modeling Hub
\citep{FluSMH}. This format includes multiple simulated values (samples)
from the forecast distribution. Support for the `sample' output type is
under consideration for future releases, and, in general, extensions to
support more output types would aim to broaden the scope of applications
in real-world forecasting tasks.

\section*{Acknowledgements}\label{acknowledgements}
\addcontentsline{toc}{section}{Acknowledgements}

We acknowledge Zhian N. Kamvar for debugging and resolving coding issues
while developing the package. We are also grateful to Matthew Cornell
for his advice on unit testing, which greatly helped us improve the
structure and testing our code with a solid understanding of unit
testing practices. Moreover, we would like to thank the hubverse
development team for their data standards, on which our package is
based.

\section*{Appendix}\label{appendix}
\addcontentsline{toc}{section}{Appendix}

\appendix

\section{Weights for subsets in LASOMO}\label{ch2sec:appendix}

In the LASOMO algorithm, two weighting schemes are available for subsets
of models in the calculation of model importance scores: equal weights
and permutation-based weights.

Let \(n\) be the total number of models and \(k\) be the size of a
subset that does not include the model being evaluated. The formulas for
the weights under each scheme are as follows: \begin{align*}
w^{\text{eq}} &= \frac{1}{2^{n-1}-1}, \\
w^{\text{perm}} &= \frac{1}{(n-1)\binom{n-1}{k}},
\end{align*} where the superscripts ``eq'' and ``perm'' denote the equal
and permutation-based weighting schemes, respectively. The maximum
weight under the permutation-based scheme occurs when \(k=n-1\), which
is \({1}/{(n-1)}\). The minimum weight occurs when the subset size is
around \({(n-1)}/{2}\) (i.e., \(k=\lfloor (n-1)/2 \rfloor\)), which is
approximately \(\displaystyle\frac{\sqrt{\pi(n-1)/2}}{(n-1)2^{n-1}}\) by
Stirling's approximation.

Given a fixed mid-sized subset, as \(n\) increases, the weight assigned
to this subset under the equal weighting scheme decreases at a rate of
\(O({1}/{2^n})\), while under the permutation-based scheme, it decreases
at a much faster rate of \(O({1}/({\sqrt{n}\,2^n}))\). This indicates
that as the number of models grows, that mid-sized subset becomes
significantly less influential in determining model importance scores
when using the permutation-based weighting scheme compared to the equal
weighting scheme.

On the other hand, for subsets of extreme sizes (e.g., \(k=1\) or
\(k=n-1\)), the weights under permutation-based weighting scheme
decrease only at \(O({1}/{n})\), much slower under the equal weighting
scheme. This implies that in scenarios with a large number of models,
the contributions of these extreme-sized subsets play a relatively
larger role in the calculation of model importance scores when using
permutation-based weights compared to the equal weighting approach.

\bibliographystyle{plain}
\bibliography{references}
\end{document}